\documentclass[prl,twocolumn]{revtex4}

\usepackage{graphicx}
\usepackage{subfigure}

\newcommand{\nv}{NV$^-$}
\newcommand{\nvsp}{NV$^-$ }
\newcommand{\nvs}{NV$^-$s}

\begin{document}

%\preprint{}

\title{Strongly enhanced photon collection from diamond defect centres under micro-fabricated integrated solid immersion lenses} %Title of paper


\author{J. P. Hadden}
\email[]{JP.Hadden@bristol.ac.uk}
%\thanks{}
%\affiliation{University of Bristol}
%

\author{J. P. Harrison}
\author{A. C. Stanley-Clarke}
\author{L. Marseglia}
\author{Y-L. D. Ho}
\author{B. R. Patton}
\author{J. L. O'Brien}
\author{J. G. Rarity}
\affiliation{Centre for Quantum Photonics, H. H. Wills Physics Laboratory \& Department of Electrical and Electronic Engineering, University of Bristol, Merchant Venturers Building, Woodland Road, Bristol, BS8 1UB, UK}

\date{\today}

\begin{abstract}
The efficiency of collecting photons from optically active defect centres in bulk diamond is greatly reduced by refraction and reflection at the diamond-air interface. We report on the fabrication and measurement of a geometrical solution to the problem; integrated solid immersion lenses (SILs) etched directly into the surface of diamond.  An increase of a factor of 10 was observed in the saturated count-rate from a single negatively charged nitrogen-vacancy (\nv) within a 5 $\mu$m diameter SIL compared with \nv s under a planar surface in the same crystal.  A factor of 3 reduction in background emission was also observed due to the reduced excitation volume with a SIL present.   Such a system is potentially scalable and easily adaptable to other defect centres in bulk diamond.
\end{abstract}


\maketitle %\maketitle must follow title, authors, abstract and \pacs

%\section{}
%%\label{}
%\subsection{}
%\subsubsection{}

The ability to address single defect centres in diamond using confocal microscopy allows optical access to these single `atom like' systems trapped within a macro-scale solid.  The negatively charged nitrogen-vacancy centre (\nv) is of particular interest for applications such as single photon generation\cite{Kurtsiefer2000,Beveratos2002}, nanoscale magnetometery\cite{Balasubramanian2008}, and fundamental investigations of spin interactions and entanglement at room temperature\cite{Jelezko2004,Hanson2006,Neumann2008}.  Other defect centres that exhibit single photon emission have also been identified (e.g. the nickel-related `NE8'\cite{Gaebel2004}, the silicon-vacancy\cite{Wang2006}, and chromium related centres\cite{Aharonovich2010}), but the search continues for other defect centres with spin properties like those of the \nvsp centre\cite{Weber2010}.  The high refractive index of diamond causes refraction of the emitted light at the diamond-air interface, reducing the possible angular collection of a microscope objective.  Thus the \nvsp photon collection efficiency is severely reduced.  This is a problem regardless of the application, or of the particular defect centre of interest.  Here we report on the fabrication and measurement of hemispherical integrated solid immersion lenses (SILs) etched directly into the surface of diamond. These structures eliminate surface refraction, thus increasing the numerical aperture (NA) of the microscope system.  This allows a substantial increase in the resolution and background rejection of our system, along with a strong enhancement in \nvsp photon collection efficiency.  Moreover,
this geometrical solution can easily be applied to other defect centres in bulk diamond which emit at different wavelengths\cite{Liu2005}.

The photon collection efficiency from \nv centres in diamond has previously been improved by using \nvsp centres located within nanocrystals small enough that the centres effectively emit into free space\cite{Beveratos2002,Ampem-Lassen2009}, or nanophotonic structures such as nanowires which guide emission towards collection\cite{Babinec2010}.  Photon collection is increased by a factor of up to about 5 in the former case and 10 in the latter.  However with the \nvsp centres positioned so close to the surface, local strain, impurities and other surface effects have been shown to degrade the stability, and spin coherence time of the \nvsp centre\cite{Bradac2010,Rabeau2007}, so a solution which improves the photon collection efficiency from \nvsp centres in bulk diamond is desirable.  By etching hemispherical SILs into the surface of the diamond we can increase photon collection efficiency without requiring the centre to be close to the surface.  Rays traced from a defect centre located at the origin of a hemispherical diamond surface are normal to the surface at all points on the hemisphere.  Therefore no refraction occurs and the theoretical NA of collection is increased by a value equal to the refractive index ($n_{d}$ = 2.42) of diamond\cite{Corle1996}.  This allows a significant increase in the collection efficiency.

Other advantages related to the increased NA of a SIL are an imaging magnification (by a factor of $n_{d}$ laterally) and a reduced (by $n_{d}^2$ longitudinally) laser excitation volume\cite{Corle1996}. The latter both increases the resolution and boosts the effective power density for a given input laser power.  SILs (also known as numerical aperture increasing lenses) have previously been used to boost photon collection from quantum dots\cite{Zwiller2001,Liu2005} and single molecules in anthracene crystals\cite{Trebbia2009}.  Both implementations used millimetre scale SILs placed on the surface of the sample. These are less scalable and also incur losses and aberrations caused by SIL-sample mismatch, surface reflections and gaps between the surfaces.  The use of a micro-scale integrated SIL overcomes these problems.  Integrated micro-lenses have previously been fabricated in diamond using inductively-coupled plasma etching (and other etching techniques), however they have not been used to improve collection of defect centre emission\cite{Lee2008}.

\begin{figure}
    \includegraphics{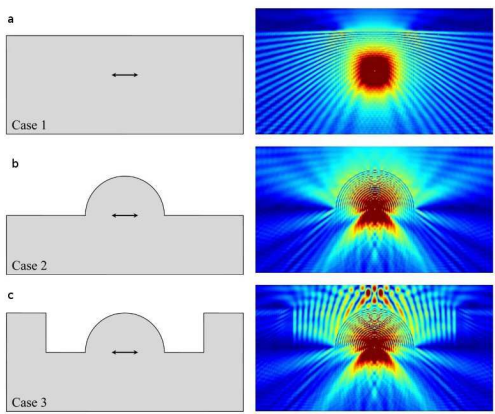}
    \caption{\label{figFDTD} Representation of electromagnetic field intensity in the YZ plane for a single frequency calculated from FDTD simulations for: (a) dipole 2.5 $\mu$m beneath a planar diamond surface, (b) dipole at origin of 2.5 $\mu$m radius hemisphere, (c) dipole at origin of 2.5 $\mu$m radius hemisphere with 2 $\mu$m wide trench.}
\end{figure}

Using a finite-difference time-domain (FDTD) method, we simulated the collection efficiency (into a microscope objective with 0.9 NA) for a dipole located: 2.5 $\mu$m below a planar diamond surface; at the focal point of a 2.5 $\mu$m radius hemisphere; and at the focal point of a 2.5 $\mu$m radius hemisphere surrounded by a 2 $\mu$m wide trench.  The latter case is included because to etch the entire surface of the sample to the base of the SIL, as in the ideal case, would be too time consuming.  The size of the trench is chosen such that the entire numerical aperture of the collection lens is utilised.  It can be clearly seen from a cross-section of the simulated electromagnetic field intensity shown in Fig. \ref{figFDTD} that the SIL increases the field intensity crossing the diamond-air interface.

The collection efficiencies were calculated for wavelengths in the range of 600-800 nm (covering the majority of the \nvsp emission spectrum) and then an average was taken. The collection efficiencies calculated in this way were 5.6\%, 29.8\%, and 28.6\% respectively. The first two cases are consistent with efficiencies calculated by purely analytic methods\cite{Barnes2002}. In other words, we expect a $\sim$5-fold increase in collection efficiency from geometrical considerations alone.  However it should be noted that in the planar case the effect of spherical aberration, which is significant for diamond material, is not included in the FDTD simulations or the analytic model.  Spherical aberrations are eliminated in the hemispherical SIL geometry, therefore we expect the measured enhancement in photon collection efficiency to be greater than 5.

The FDTD simulations also allow us to investigate the performance of the SILs when the dipole is not located precisely at the origin of the hemisphere. Simulations show that if the dipole position is varied by 1 $\mu$m  along any of the three Cartesian axes the collection efficiency should remain above ~20\% (c.f. 28.6\% at the focus).  In other words the SILs are relatively tolerant to lateral or longitudinal placement error.  It should also be noted that the magnification effect of the SILs means that a 1 $\mu$m lateral displacement in the diamond corresponds to a measured `image' distance (in a confocal scan) of 2.42 $\mu$m, placing it at the edge of the lens in a confocal image.

\begin{figure}
 \subfigure{\label{figSILsFIB}}
 \includegraphics[width=8.5cm]{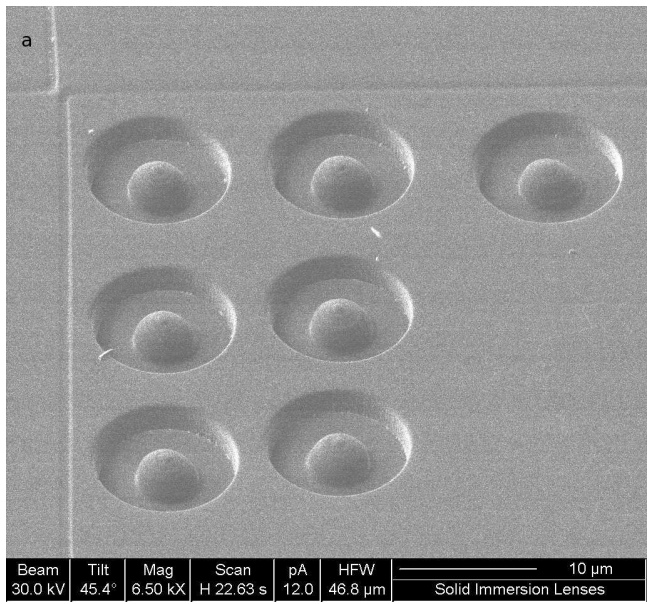}
  \subfigure{\label{figSILsConfocal}}
 \includegraphics[width=8.5cm]{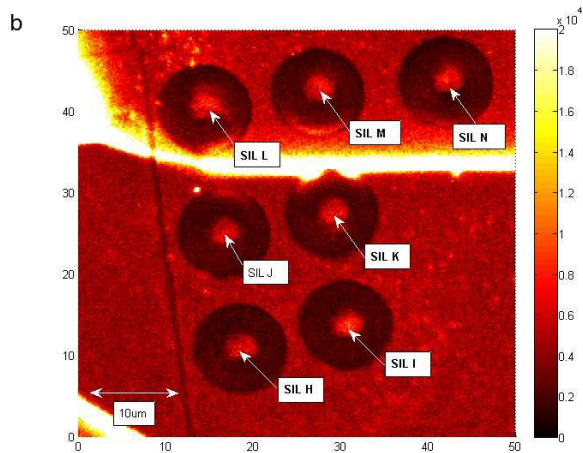}
 \caption{\subref{figSILsFIB} FIB image showing 7 SILs and location grid lines. \subref{figSILsConfocal} Confocal image of the same area with SILs labelled.  The bright line is a diamond crystal grain boundary.}
\end{figure}

\begin{figure*}
     \includegraphics[width=14cm]{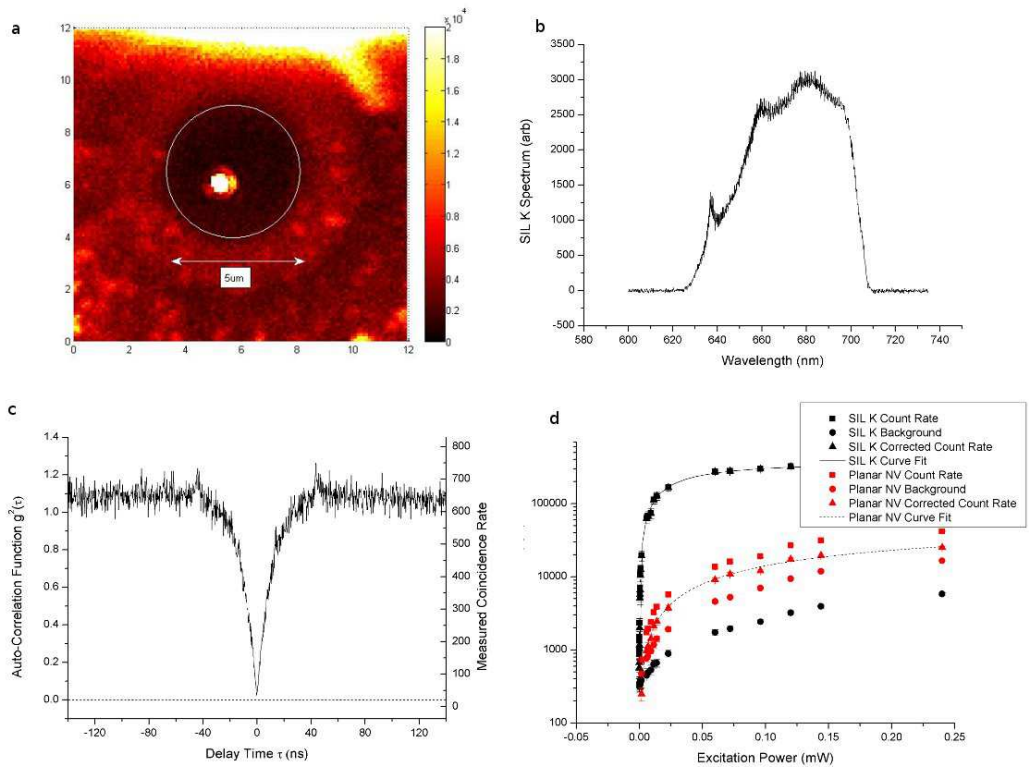}
        \label{AllFigures}\subfigure{\label{figConfocal}}\subfigure{\label{figSpectrum}}\subfigure{\label{figAB}}\subfigure{\label{figSaturation}}
        \caption{\subref{figConfocal} Confocal image of SIL K.  The boundary of the SIL is marked in white.  Note that the colour scale has been chosen to maximise the visibility of the SIL and surroundings - the intensity of the bright spot is about 120 kcps.
        \subref{figSpectrum} Optical spectrum recorded at the point of highest intensity in \subref{figConfocal}.  The sharp drops in spectral intensity at 630 nm and 700 nm are caused by the spectral filter and not by chromatism in the SIL.
        \subref{figAB} Second order intensity correlation function recorded at the same point as in \subref{figSpectrum}.
        \subref{figSaturation} Comparison of the total count rate for the single \nvsp under SIL K with that for a single \nvsp under a planar surface as a function of excitation power.  The background in each case is measured from a point close to the \nvsp centre.}
\end{figure*}

SILs were fabricated in a polycrystalline type IIa CVD diamond sample (Element Six) measuring $\sim$5 mm $\times$ 5 mm and of 1 mm thickness. There is a large variation in the density of \nvsp centres in the sample, and a region was chosen where the density was low enough to resolve single \nvs.  A 30 keV gallium focused ion beam system (FEI Strata FIB-201) was used to fabricate the SILs.  Hemispheres were approximated by etching concentric rings of increasing depth and diameter.  This was achieved by varying the beam current and dwell time and water was used throughout to assist the etching process.  A FIB image showing 7 of the 14 SILs fabricated along with FIB etched grid lines is shown in Fig. \ref{figSILsFIB}, and a confocal image of the same area with the SILs labelled is shown in Fig. \ref{figSILsConfocal}.

Optical characterisation was performed using a laser scanning confocal microscope system built in-house. The output from a frequency-doubled Nd:YAG laser (Cobolt Samba) was focused onto the sample using a 0.9 NA microscope objective (Nikon). Laser-induced fluorescence was collected by the same objective and focused onto the 8.9 $\mu$m core of an optical fibre (serving as the confocal aperture). A 532 nm long-pass filter was used to reject stray laser light. A band-pass filter centred at 675 nm with a 67 nm transmission band was used to block first and second order Raman scatter.  It was tilted to shift the transmission band to 630-700 nm so that the \nvsp zero-phonon emission at 637 nm was transmitted. The fluorescence photons were then directed either to a spectrograph (Shamrock 303) fitted with a CCD camera (Newton 920) to record optical spectra, or to a Hanbury-Brown and Twiss detection scheme to record the second order intensity correlation function ($g^{(2)}(\tau)$).

Of the 14 SILs etched, 2 fell on grain boundaries and were not investigated further. Of the remaining 12, 5 contained single \nvsp centres. One of these 5 actually had two centres that could be resolved and analysed separately.  Only  data from the `best' of the single \nvsp containing SILs (denoted `SIL K') is presented in full here.  The enhancement of the other SILs is commented on briefly.

A confocal scan of a 20 $\times$ 20 $\mu$m$^2$ area of the sample in the region of SIL K is shown in Fig. \ref{figConfocal}.  A high intensity region is evident near the origin of the SIL.  This region is 1 $\mu$m away from the SIL's origin in the confocal scan, corresponding to 0.41 $\mu$m in `real' space.  The bright feature at the top is a grain boundary.

The optical spectrum recorded at this point (Fig. \ref{figSpectrum}) displays the characteristic \nvsp emission profile, with zero-phonon line around 637 nm and broad phonon assisted side-band at longer wavelengths.  The anti-bunching dip in the second order intensity correlation function (Fig. \ref{figAB}) clearly indicates that the emission arises from a single centre.  The data has been normalised and corrected for background as described by Beveratos \emph{et al.}\cite{Beveratos2002}.

To quantify any enhancement in collection efficiency, photon count-rate as a function of laser power was recorded for the \nvsp in SIL K and for other single \nvsp centres in the same diamond grain under an unmodified planar surface.  A comparison of the saturated intensities (Fig. \ref{figSaturation}) indicates an enhancement of $\sim$10, with a measured absolute count rate of 345 kcps.  This compares favourably with the enhancement and absolute count rates reported from diamond nanocrystals\cite{Ampem-Lassen2009} and diamond nanowires \cite{Babinec2010}.  It should be noted that the absolute count rate in our system could be improved by optimising the spectral filters.  By filtering between 630 nm and 700 nm, around 70\% of the \nvsp spectrum is being collected.  In principle then, optimisation of the filters would result in an absolute count rate of ~500 kcps.

The 3 other SILs containing a single \nvsp centre had enhancement factors of: 10, 8, and 8. For the SIL with two separate single \nvs, the enhancement factors were 6 and 3.6.

\begin{figure}
 \includegraphics[width=7.5cm]{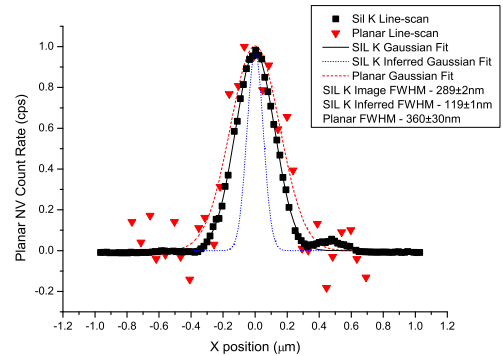}%
 \caption{\label{figLinescanTrace} Comparison of confocal line scan of the single \nvsp under SIL K and from a  single \nvsp under a planar surface both fitted with a normalised Gaussian function.  The inferred `real' fit of SIL K taking into account the magnification of the SIL is also plotted.  The full width half maximums (FWHM) of the two fits, and the inferred `real' FWHM of SIL K's are shown with their associated uncertainties.}%
\end{figure}

As well as improving the angular collection efficiency, the SILs also modify the profile of the excitation beam.  Elimination of refraction in the excitation beam means that it can be focused to a smaller volume, thus increased resolution is to be expected.  Assuming that the excitation point spread function is Gaussian in shape, we can estimate the excitation spot size from the full width half maximum (FWHM) of a line scan across an \nvsp centre.  Figure \ref{figLinescanTrace} shows a comparison between the line scan across a planar \nvsp centre and across the one under SIL K.  The FWHM is 360 $\pm$ 30 nm in the planar case and 289 $\pm$ 2 nm in the SIL case.  After compensating for the magnification of the SIL the spot size is estimated to be 120 $\pm$ 1 nm.  The theoretical single point resolution is given by FWHM = 0.37 $\lambda_{ex}$ / NA = 90 nm for our system with excitation wavelength $\lambda_{ex}$ = 532 nm and an increased NA of 2.18\cite{Corle1996}.  The reduced performance is likely to arise from aberrations within our system caused by the \nvsp centre not being precisely at the origin of the SIL.

Improved resolution is important because a smaller excitation volume decreases the proportion of unwanted background fluorescence collected along with the desired \nvsp fluorescence.  This is clearly visible in (Fig. \ref{figSaturation}) where the signal to background ratio is much higher with the SIL, and in fact the absolute value of background is 3 times lower at equivalent pump powers.  A smaller excitation volume also means that the power density is increased (for a given laser power) and less power is required to saturate an \nvsp centre.  SIL K obviously contains a very well positioned \nvsp centre and we do see variation in these enhancements from SIL to SIL.  We ascribe this not only to \nvsp position, but also to variation of dipole orientation, since the crystal orientation is unknown in this polycrystalline sample.  We are currently studying both of these effects.

This demonstration of strongly enhanced photon collection efficiency from \nvsp centres using integrated solid immersion lenses is a step towards efficient single photon sources as well as efficient optical spin read-out in compact micro-structured devices.  The enhancement compares favourably to those reported from nanocrystal and nanowire  devices\cite{Ampem-Lassen2009,Babinec2010}, with the advantage that \nvsp centres are not located close to the surface where they might suffer from surface-induced decoherence.  Since the enhancement is wavelength independent when the dipole is at the origin of the SIL it may be used with other defect centres in bulk diamond.  Such a solution is robust and potentially scalable, since SILs could be fabricated over existing defect centres after characterisation.  Refinements in surface symmetry, SIL placement and smoothness should allow us to achieve even greater enhancement as our fabrication technique is perfected.

\begin{acknowledgments}
This work was supported by EPSRC, European Union Sixth Framework Program project EQUIND IST-034368, NanoSci-ERA project NEDQIT and the Leverhulme Trust. JGR and JLOB are both supported by individual European research council fellowships and by Royal Society Wolfson research merit awards.
\end{acknowledgments}

\bibliography{bib_polysils}

\end{document}